%% file: main-mod-v2.tex
\documentclass[prd,preprint,tightenlines,floatfix,showpacs,preprintnumbers,nofootinbib,eqsecnum,superscriptaddress]{revtex4-1}
\usepackage[a4paper, top=2cm, bottom=2cm, left=1.5cm, right=1.5cm]{geometry}

\usepackage[utf8]{inputenc}
\usepackage[T1]{fontenc}
\usepackage{lmodern}
\usepackage[english]{babel}

\usepackage{color}
\usepackage{amsfonts}       
\usepackage{amsmath}        
\usepackage{amssymb}        
\usepackage{amsbsy}     
\usepackage{units}

\usepackage{soul}
\usepackage{bm}

\usepackage{graphicx}

\newcommand{\xpom}{{x_\mathbb{P}}}
\newcommand{\xP}{{x_P}}

\newcommand{\ktmin}{k_{\perp\mathrm{min}}}
\newcommand{\beq}{\begin{align}}
\newcommand{\eeq}{\end{align}}

\newcommand{\LQCD}{\Lambda_\mathrm{QCD}}
\newcommand{\ktminZ}{k_{\perp0}}

\newcommand{\LEPTO}{\sc Lepto}
\newcommand{\PYTHIA}{\sc Pythia}

\newcommand{\CASCADE}{\sc Cascade}

\newcommand{\sFWD}{\sigma^D_{r,\mathrm{FWD}}}
\newcommand{\sLRG}{\sigma^D_{r,\mathrm{LRG}}}

\begin{document}

\begin{flushright}
LU TP 15-51\\ 
April 2016 
\vskip1.5cm
\end{flushright}

\title{Dynamic colour screening in diffractive deep inelastic scattering}

\author{Gunnar Ingelman}
\email{Gunnar.Ingelman@physics.uu.se}
\affiliation{Department of Physics and Astronomy,
Uppsala University, Box 516, 75120 Uppsala, Sweden}

\author{Roman Pasechnik}
\email{Roman.Pasechnik@thep.lu.se}
\affiliation{Department of Astronomy and Theoretical Physics,
Lund University, 22362 Lund, Sweden\vspace{1cm}}

\author{Dominik Werder}
\email{Dominik.Werder@physics.uu.se}
\affiliation{Department of Physics and Astronomy,
Uppsala University, Box 516, 75120 Uppsala, Sweden}

\begin{abstract}
\vspace{0.5cm}
We present a novel Monte-Carlo implementation of dynamic colour screening via multiple exchanges of semi-soft gluons as a basic QCD mechanism 
to understand diffractive electron-proton scattering at the HERA collider. Based on the kinematics of individual events in the standard QCD description 
of deep inelastic scattering at the parton level, which at low $x$ is dominantly gluon-initiated, the probability is evaluated for additional exchanges 
of softer gluons resulting in an overall colour singlet exchange leading to a forward proton and a rapidity gap as the characteristic observables for diffractive scattering.
The probability depends on the impact parameter of the soft exchanges and varies with the transverse size of the hard scattering subsystem and is therefore 
influenced by different QCD effects. We account for matrix elements and parton shower evolution either via conventional DGLAP $\log Q^2$-evolution 
with collinear factorisation or CCFM small-$x$ evolution with $k_\perp$-factorisation and discuss the sensitivity to the gluon density distribution 
in the proton and the importance of large log\,$x$-contributions. The overall result is that, with only two model parameters which have theoretically 
motivated values, a satisfactory description of the observed diffractive cross-section at HERA is obtained in a wide kinematical range.
\end{abstract}

\maketitle

\section{Introduction}

Diffractive scattering through strong interactions without any large momentum
transfer has historically been described in terms of the exchange of a Pomeron,
a virtual hadron-like object with vacuum quantum numbers. The Regge
approach~\cite{Polkinghorne:1980mk,Forshaw:1997dc} of the pre-QCD era provides
a working phenomenology to describe such processes in a hadron basis since no
parton structure is resolved. The idea \cite{Ingelman:1984ns} to introduce a
hard scale in a diffractive process opened the possibility to examine these
processes at the level of quarks and gluons in the modern framework of QCD. The
discovery of such {\em diffractive hard scattering} was made by the UA8
experiment \cite{Bonino:1988ae} by observing high-$p_\perp$ jets in single diffractive
events in $p\bar{p}$ collisions at CERN. Many other hard processes in
diffractive events have been observed later on, for a review see e.g.\
\cite{Ingelman:2005ku}. Of special importance are the measurements of diffractive deep
inelastic scattering (DDIS) at the electron-proton collider HERA
\cite{DDIS-exp,DDIS-latest}, where the well understood point-like
electromagnetic interaction probes the parton structure in the diffractive
reaction mechanism. 

This has resulted in theoretical descriptions of data based on essentially two
different approaches, one being pomeron exchange in Regge phenomenology and the
other colour screening via soft gluon exchange in QCD. 

The pomeron approach starts with the initial proton state fluctuating in a soft
non-perturbative process into a proton and a pomeron. The latter is assumed to
have a partonic structure which is probed by the momentum transfer $Q^2$ of the
deep inelastic photon exchange to produce the hadronic final state, well
separated in rapidity from the final state proton carrying most of the
longitudinal momentum of the beam proton. HERA data on DDIS can then be
described \cite{Ingelman:1984ns, Ingelman:1992qf} as a product of an effective
pomeron flux factor from Regge phenomenology and deep inelastic scattering on
the pomeron having parametrised parton density functions (PDF) \cite{Ingelman:1992qf}.
Alternatively, one may parametrise diffractive structure functions without an
explicit pomeron flux but instead being conditionally dependent on the momentum
of the final proton.

However, this approach is not universal in the sense that such parametrisations
do not reproduce diffractive hard scattering data in hadron-hadron
collisions. For example, such pomeron PDFs overestimate substantially the cross-sections for
diffractive hard scattering processes, such as production of jets or $W$, at
the Tevatron \cite{Affolder:2000vb}.
This has called for introducing a gap survival suppression factor $\hat S$,
which can be given a qualitative theoretical motivation but which is difficult
to calculate quantitatively.

The colour exchange approach starts instead with the hard scattering process
and then adds softer gluon exchanges to achieve the effective colour singlet
exchange in diffractive processes. Thus, the underlying hard process is assumed
to be the same as in the corresponding non-diffractive process and its momenta
naturally not affected by other exchanges at much lower momentum transfer
scales. However, the formation of confining colour fields may well be affected
by the softer gluon exchanges and thereby the hadronisation process such that a
different distribution of the final state hadrons emerges. For example, when
different colour singlet string-fields emerge separated in rapidity they will
hadronise into two hadron systems separated by a rapidity gap with no hadrons. 

A simple, but phenomenologically rather successful model of this kind is the
Soft Colour Interaction (SCI) model~\cite{Edin:1995gi} added to the {\LEPTO}
\cite{Ingelman:1996mq} and {\PYTHIA} \cite{Sjostrand:2006za} Monte-Carlo event
generators. A large variety of diffractive data could then be reproduced with
essentially the same value of a single new parameter, introduced to give the
probability for exchange a soft colour octet gluon between any pair of partons.
This colour exchange alter the formation of the colour string-fields and hence
the application of the conventional Lund string hadronisation model results in
a different topology of the hadronic final state. The model gives an
essentially correct description of diffractive DIS $ep$ scattering at HERA
\cite{Edin:1995gi} as well as diffractive events at the Tevatron having jets or
quarkonia \cite{Edin:1997zb,Enberg:2001vq} or gauge bosons
\cite{Ingelman:2012wj}. It has therefore been applied for predictions, e.g. of
diffractive Higgs production in double gap processes at LHC
\cite{Enberg:2002id}. The gap survival factor $\hat S$ often used in other
kinds of models, is not necessary here since the full event simulation accounts
for such effects resulting in correct rates for the investigated diffractive
processes in $ep$ and $pp$ and $p\bar p$ collisions.

Other models of a similar nature, with different forms of colour exchange, have
been developed. The GAL model \cite{Rathsman:1998tp} for example considers soft colour
exchanges between strings with a probability that favours minimization of the
strings' area in energy-momentum coordinates. A recent development of such
colour string reconnection models makes a more elaborate account for $SU(3)$
colour statistics \cite{Christiansen:2015yqa}.

A theoretical QCD basis for basic colour exchanges has been proposed in
\cite{Brodsky:2004hi} and, in a more elaborated form, in
\cite{Pasechnik:2010zs}. The basic hard scattering process is treated by
conventional perturbative QCD (pQCD). Its large momentum scale implies that it
occurs on a small space-time scale compared to the bound state proton and is
thus embedded in the proton. Therefore, the emerging hard-scattered partons
propagate through the proton's colour field and may interact with it. The
amplitude for such multiple gluon exchanges is calculated in the eikonal
approximation to all orders in perturbation theory resulting in an analytic
expression for such a colour screening effect. The theoretical approach 
\cite{Pasechnik:2010zs} used here to resum multigluon exchange has similarities 
with the one in \cite{HKS,HS} giving a similar eikonal factor $1 - \exp(\dots)$ 
in the amplitude, but there are also differences as will be discussed below. 
It is this amplitude that we here develop into a probabilistic model. We implement it for 
two different Monte Carlo event generators: {\LEPTO} for general deep inelastic
lepton-nucleon scattering including first order QCD matrix elements and parton
showers based on conventional $\log{Q^2}$ evolution from the DGLAP equations
\cite{DGLAP}, and as well for {\CASCADE} \cite{CASCADE} specialised on small-$x$
electron-proton scattering based on the off-shell $\gamma^\star g^\star\to
q\bar{q}$ first order matrix element and $k_\perp$ factorisation with CCFM
evolution \cite{CCFM} and unintegrated gluon density of the proton. 

The paper is organized as follows. In section \ref{sec:dynamic-rescattering} we
describe the basic process with the colour screening. Section
\ref{sec:ddis-via-dcs} discusses the resulting cross-section for
diffractive deep inelastic scattering and its Monte Carlo implementation.
Section \ref{sec:res} shows our results in comparison to HERA data.  Finally, section
\ref{sec:conclusions} presents our conclusions.

\section{Soft colour screening in diffractive DIS}
\label{sec:dynamic-rescattering}

\begin{figure}[!h]
\begin{center}
\includegraphics[width=0.5\textwidth]{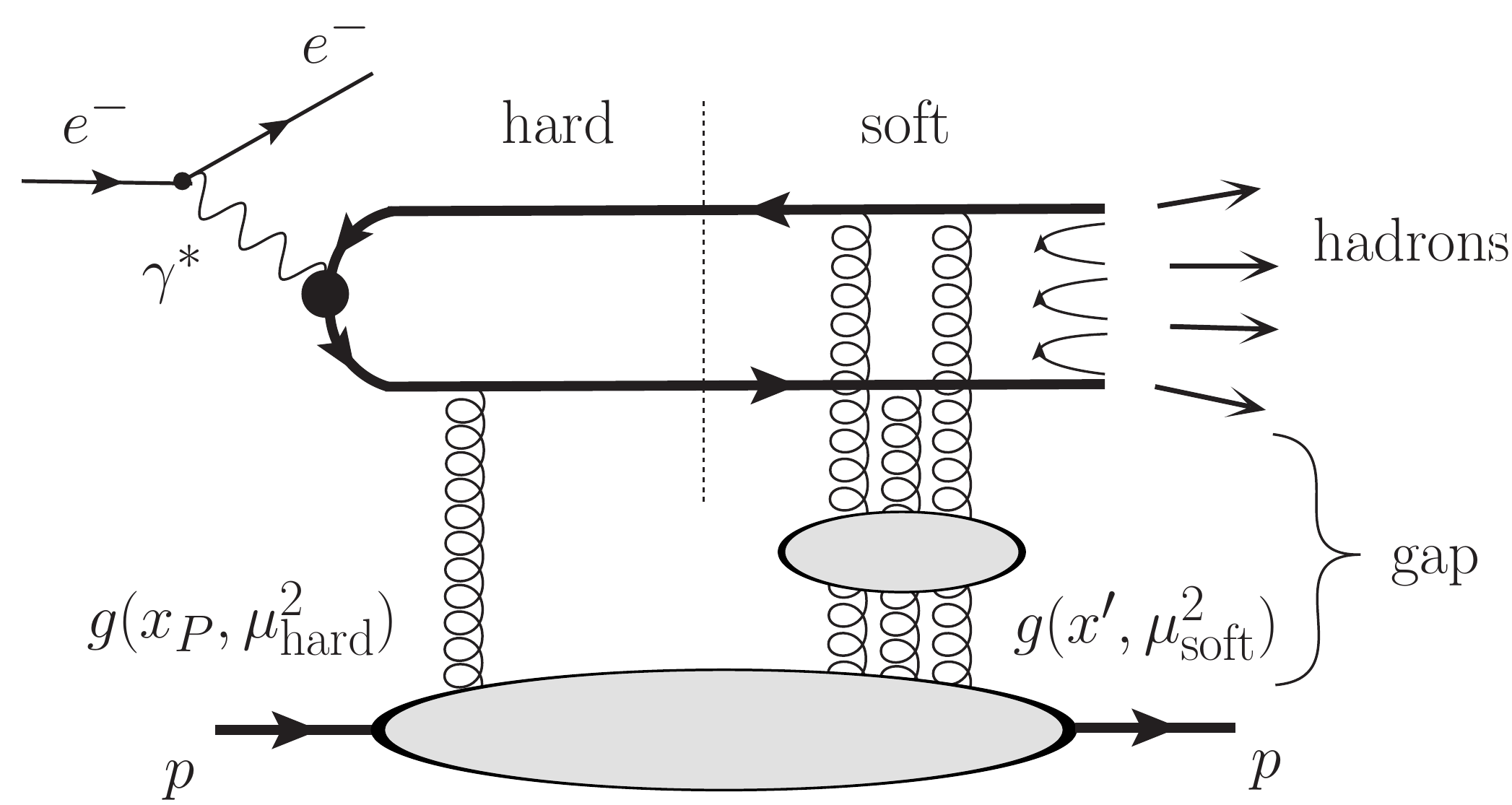}
\includegraphics[width=0.35\textwidth]{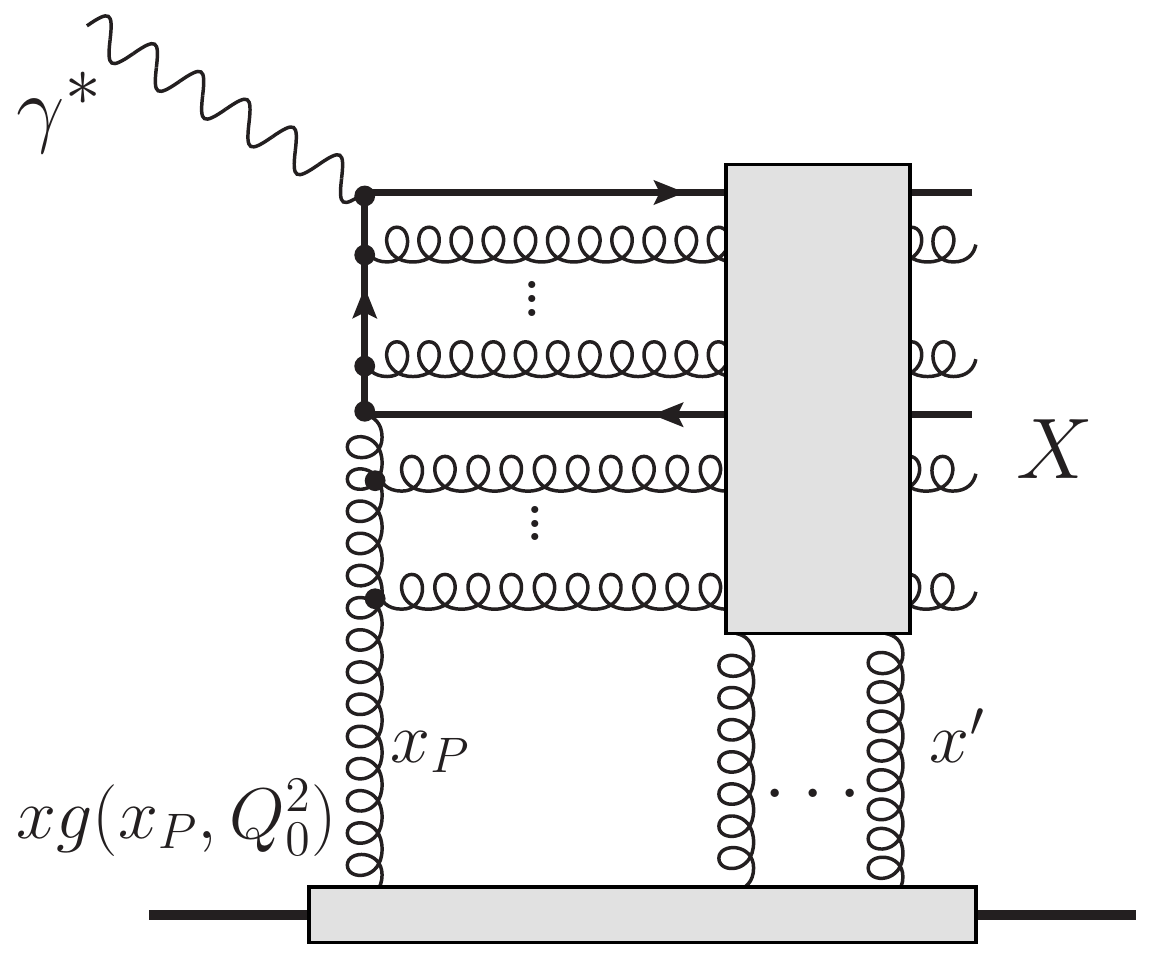}
\caption{Illustration of the diffractive DIS process with the hard subprocess matrix 
element $\gamma^*g\to q\bar q$ with a subsequent rescattering of the $q\bar q$ dipole
off the target colour field (left panel). Schematic diagram of the diffractive DIS process 
$\gamma^*p \to X+{\rm gap}+p$ accounting for final state rescattering by multiple gluon 
exchange at $x'\ll \xpom$ and perturbative parton shower off initial state parton which 
builds up the diffractive system $X$ (right panel). The latter can be separated from 
the leading proton (or small-mass system $Y$) by a rapidity gap. The final state radiation 
is not shown as it does not affect the overall kinematics of the $X$ system.}
\label{fig:DDIS}
\end{center}
\end{figure}

In the virtue of the SCI model, the skeleton of both inclusive and diffractive (with rapidity gaps and/or leading proton) 
DIS process is provided by the same perturbative QCD diagram illustrated in Fig.~\ref{fig:DDIS}. A parton with longitudinal 
momentum fraction $x_P$ in the initial proton at the starting scale $Q_0^2$ for pQCD is evolved to smaller momentum 
fractions, but higher transverse momenta and virtualities up to a hard scale $\mu_{\rm hard}^2\simeq Q^2$. Here, 
a virtual photon $\gamma^*$ with momentum $q$ and virtuality $Q^2=-q^2$ resolves a quark at Bjorken $x$
\begin{eqnarray}
\label{Bjx}
x = \frac{Q^2}{2 P \cdot q} = \frac{Q^2}{Q^2 + W^2} \,.
\end{eqnarray}
At small $x$ the process will dominantly be initiated by a gluon, which can radiate and  splits into a $q\bar{q}$ pair. 
The total momentum fraction taken from the proton is $\xP = x/\beta$, and the total mass $M_X$ of the parton system denoted
as $X$ in Fig.~\ref{fig:DDIS} is
\begin{eqnarray}
\label{MXdef}
M_X^2 = Q^2 \left(\frac{1}{\beta} - 1\right) = Q^2 \left(\frac{\xpom}{x} - 1\right) \,.
\end{eqnarray}
For a $\gamma g\to q\bar q$ pair (without additional gluons) having with quark transverse momentum $k_\perp$ and 
longitudinal momentum fraction $z$, the corresponding invariant mass is
\begin{eqnarray}
\label{MXdef-qq}
M_X^2 = \frac{k_\perp^2 + m_q^2}{z(1-z)} \,.
\end{eqnarray}

At large $\xP$, the parton distribution functions (PDFs) of the incident proton
are dominated by valence quarks leaving practically no chance for the proton
to survive such an interaction, and hence resulting in a non-diffractive event.
At small $\xP$, however, the PDFs are dominated by gluons, and the partonic system $X$
is created in photon-gluon fusion $\gamma^*+g\to q\bar{q}$ as depicted in Fig.~\ref{fig:DDIS}. In this case,
the momentum exchange via multiple soft gluons with a small net fraction $x' \ll \xP$
between the proton and the perturbative $X$ systems does not significantly change the momenta of
partons emerging from the hard scattering. However, they do change the colour structure 
of the resulting $X$ and $Y$ systems.

The original SCI model captures the main effects in many processes
as DDIS at HERA and diffractive hard scattering at the Tevatron.
Despite this success, it is not derived from a perturbative QCD amplitude.
For the case of DDIS, a derivation of the amplitude for the colour screened
process has been done based on perturbative QCD in the large $N_C$ limit
\cite{Brodsky:2004hi, Pasechnik:2010zs}
and provides a theoretical basis for the DDIS process in terms of colour exchanges.
It improves on the previous description by introducing a dependence on the kinematical
details of the event which also leads to colour transparency.
We start with the outline of the resummed colour screening amplitude and
the derivation of probability for an event-by-event treatment in an event generator.

Consider the DDIS amplitude in impact parameter representation in the target rest frame which corresponds to
the colour dipole picture of the process. The lowest Fock component of the virtual photon $\gamma^* \to q\bar{q}$
corresponds to its fluctuation to a $q\bar q$ dipole with transverse separation ${\bf r}$ in the colour background field
of the traget proton at impact distance ${\bf b}$ from its centre. The prepared Fock component then propagates
through the field in the proton and softly interacts with it such that it can, in principle, change its colour
but not kinematics (the dipole size is frozen at the time scale of its propagation through the colour medium).
We consider the forward limit where the total transverse momentum $\bm{\delta}_\perp$ of
gluon exchanges in the $t$-channel is small,
$|\bm{\delta}_{\perp}| \equiv \sqrt{-t}\simeq \mu_{\rm soft}\sim \Lambda_{\rm QCD}$.
In this limit, as a straightforward consequence of the optical theorem in the limit of large
$\gamma^*p$ c.m.\ energy, the DDIS amplitude $M_\mathrm{diff}$ can be written with the
ordinary gluon-initiated inclusive DIS amplitude $M_g \equiv M_{g+\gamma \to X}$
and the dynamical colour screening (DCS) amplitude $\mathcal{A}_\mathrm{DCS}$ as
\begin{equation*}
M_\mathrm{diff}(\mathbf{k}_\perp,\bm{\delta}_\perp)
\propto \int
d^2rd^2b \, M_g(\xP;\mathbf{r},\mathbf{b})
\, \mathcal{A}_\mathrm{DCS} (\mathbf{r},\mathbf{b})
\, e^{i\mathbf{r}\mathbf{k}_{\perp}} e^{i\mathbf{b}\bm{\delta}_{\perp}} \,,
\end{equation*}
where $\mathbf{k}_{\perp}$ is the relative quark transverse momentum in the $q\bar q$
dipole in the lowest order subprocess $g^*+\gamma^* \to q\bar{q}$.

The screening amplitude $\mathcal{A}_{\mathrm{DCS}}$ accounts for the
soft gluon exchanges between the proton remnant $Y$ and the rest of the final state
commonly denoted as $X$, with $X=q\bar{q}$ at the lowest order.
These exchanges carry a small longitudinal fraction $x'$ and the transverse
momentum transfer $k'_{\perp}$ is at a soft scale $\mu_\mathrm{soft}$.
$\mathcal{A}_{\mathrm{DCS}}$ is resummed to all orders
in the large-$N_c$ limit where it acquires a simple eikonal form
\cite{Brodsky:2004hi,Pasechnik:2010zs}.
\begin{equation}
\mathcal{A}_\mathrm{DCS}(\mathbf{r},\mathbf{b})
= 1 - \exp\Big(i C_F \alpha_s^\mathrm{eff}(\mu^2_\mathrm{soft})
\, \ln\frac{|\mathbf{b}-\mathbf{r}|}{|\mathbf{b}|}\Big) \,.
\label{eik-formula}
\end{equation}
Here, $C_F\simeq T_FN_c$ is the colour factor for the single gluon exchange amplitude
and $\alpha_s^\mathrm{eff}$ is the effective coupling constant
at the soft hadronic scale $\mu_\mathrm{soft}$.
The effective QCD coupling is not small in this case.
Several approaches dealing with the Landau singularities at low momentum transfers
were proposed in the literature, e.g.~\cite{Brodsky:2010ur}.
In practice, we use the infrared-stable Analytic Perturbation Theory (APT)
approach~\cite{Shirkov:1997wi}.

For our study we use similarly the inclusive amplitude
\begin{equation}
M_\mathrm{incl}(\mathbf{k}_\perp,\bm{\delta}_\perp)
\propto \int
d^2rd^2b \, M_g(\xpom;\mathbf{r},\mathbf{b})
\, e^{i\mathbf{r}\mathbf{k}_{\perp}} e^{i\mathbf{b}\bm{\delta}_\perp}
\,.
\end{equation}
In impact parameter space the fraction of the cross section with colour screening
between the systems $X$ and $Y$ is obtained from the ratio
\begin{align}
\frac{|M_\textrm{diff}(\mathbf{r},\mathbf{b})|^2}{|M_\textrm{incl}(\mathbf{r},\mathbf{b})|^2}
= |\mathcal{A}_\textrm{DCS}(\mathbf{r},\mathbf{b})|^2
\equiv P(\mathbf{r},\mathbf{b})
\label{eq:prob}
\end{align}
which defines the probability function for the overall colour singlet exchange. 
With $\mathbf{r}\mathbf{b} \equiv r b \cos\varphi$, this leads to
\begin{equation}
P(r/b,\varphi) = \left| 1-\exp
	\left(
		i C_F \alpha_s \ln \sqrt{1+\frac{r^2}{b^2}-2\frac{r}{b}\cos\varphi}
	\right)
\right|^2 \;.
\label{eq:prob-from-norm-and-angle}
\end{equation}
To apply this as the probability in a Monte-Carlo generated event at the parton level, we will associate
$r \simeq \mathbf{k}_\perp^{-1}$ and
$b \simeq \bm{\delta}_\perp^{-1}$,
and approximate by taking the average over the relative angle $\varphi$
\begin{equation}
P(r/b) = \int \frac{d\varphi}{2\pi} P(r/b,\varphi)
\label{eq:prob-from-rb}
\end{equation}
which is motivated by the fact that $\varphi$ will be uniformly distributed
in a sample of many collisions.
\begin{figure}[!h]
\begin{minipage}{0.475\linewidth}
\includegraphics[width=1\linewidth]{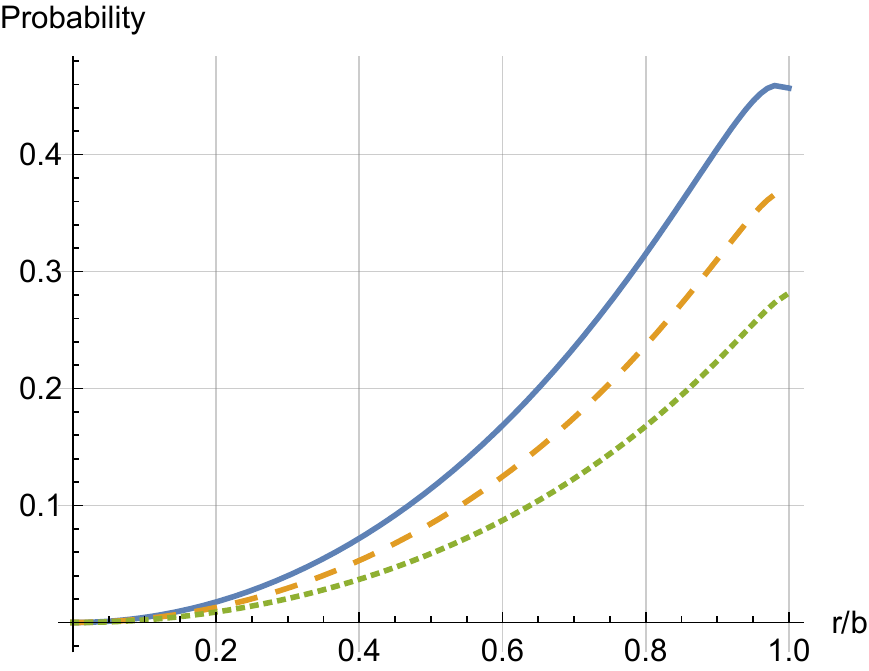}
\end{minipage}
\caption{
The screening probability $P(r/b)$ for different values of
$\alpha_s^{\rm eff} \in \{0.7, 0.6, 0.5\}$ (upper to lower curve)
}
\label{fig:rescatter-prob-for-angles}
\end{figure}

The resulting colour screening probability is shown in Fig.~\ref{fig:rescatter-prob-for-angles}, for different choices 
of $\alpha_s^\mathrm{eff}$ which enters as a normalisation factor. Two important characteristic properties can 
here be observed. First, infrared safety with the probability levelling off at large $r/b$, which resembles the saturation 
feature of the dipole scattering amplitude. Second, vanishing colour screening probability for small dipoles $r/b \ll 1$, 
which is compatible with the colour transparency property. 

Our basic theoretical approach \cite{Pasechnik:2010zs} for resumming multiple gluon exchanges has similarities with the theory 
developed in \cite{HKS,HS} and resulting in an amplitude with an eikonal factor of the form $1 - \exp(\dots)$ in similarity with 
Eq.~(\ref{eik-formula}) above. That approach can also be applied to both inclusive and diffractive DIS starting from the same 
hard matrix element and employing factorization and resummation of soft $t$-channel gluons. Despite these similarities, 
there are significant differences. In \cite{HKS,HS} all exchanged gluons are treated on the same footing via a Wilson line 
resumming them into a color singlet exchange. Our approach separates one ``leading'' gluon, carrying the largest $x$-fraction, 
from the rest of the exchanged gluons which are much softer, carrying $x’ \ll \xP$. These softer gluons are resummed to all 
orders and are required to be in a color octet state that matches the leading gluon to an overall color singlet exchange 
that provide the mechanism for rapidity gap formation. Our leading gluon is treated via conventional $k_\perp$-factorisation 
in terms of the unintegrated gluon distribution in the proton, thus providing a model which is explicitly the same for diffractive 
and inclusive scattering without introducing any new kind of parton distribution functions, cf Eq.~(\ref{sigmaD}) below. 
This is in contrast with the formalism in \cite{HKS,HS} which introduces diffractive parton distribution functions (PDF) 
containing non-perturbative dynamics and interpreted as the probability to find a parton with a momentum fraction $\xi$ 
in the proton, provided the proton emerges intact with momentum fraction $1-\xP$.  Such diffractive PDFs can be analyzed 
theoretically \cite{HKS,HS} which may introduce model parameters to be determined from data. 

Naturally, one should study these, and other, theoretical approaches and via data find the optimal description of data 
to understand the rather complex diffractive processes in terms of basic theory and few free parameters. Our approach 
has only two new parameters, which both have physical meanings that constrain their values to a rather narrow range. 
The following sections specify the Monte Carlo implementation of the model and show detailed numerical comparisons to data. 

\section{DDIS cross section via dynamic colour screening}
\label{sec:ddis-via-dcs}

The DDIS cross section is then obtained using the inclusive cross section and
standard inclusive parton densities together with the probability $P(r/b)$ in Eq.~(\ref{eq:prob-from-rb}) for
dynamic colour screening resulting in
\begin{eqnarray} \nonumber
\frac{d\sigma^D}{dQ^2\,d\beta\,d\xP} &=& \sum_i \iint  dx\,dr
	\, \rho(r, Q^2, \beta, \xP) \, \frac{d\hat{\sigma}}{dQ^2\,dx} \\ 
	&& f_i(x,Q^2) \, P(r/b) \, \delta(x-\xP\beta) \,, \label{sigmaD}
\end{eqnarray}
where $f_i(x,Q^2)$ are the standard inclusive parton distributions.
$\rho(r, Q^2, \beta, \xP)$ represents the differential distribution of the standard DIS
cross section in $r$ which is obtained from the parton evolution event-by-event in the Monte Carlo. 
Since $r$ represents the transverse size of the $q\bar{q}$ together with the pQCD radiation and the amplitude 
for colour screening is dominated by a rescattering off large dipoles, we use the smallest $k_\perp$ difference 
within the partonic $X$-system and let $r\simeq 1/\ktmin$. Although this $\ktmin$ is typically related to the pQCD cutoff, 
the Monte Carlo simulation can give very small such relative $k_\perp$ due to random angular orientations of the momentum 
vectors. We therefore introduce a cut-off $\ktminZ$ to avoid a spurious divergence and transverse sizes $r$ that are 
not perturbatively small. Thus, we let $r = 1/\sqrt{\ktmin^2+\ktminZ^2}$. 

The impact parameter $b\simeq 1/q_\perp$ is related to the soft transverse momentum of the screening multiple gluon 
exchange, which is expected to be well below the factorisation scale for the pQCD processes. On the other hand $q_\perp$ 
is expected to be somewhat larger than the confining energy-momentum scale $\Lambda_\mathrm{QCD}\sim \unit[200]{MeV}$ 
in order for the screening process to occur fast enough that the proton state can stay quantum mechanically coherent into the final state.
The colour screening probability therefore depends on the ratio $r/b$ given by
\begin{align}
\frac{r}{b} = \frac{q_\perp}{\sqrt{\ktmin^2 + \ktminZ^2}}\,,
\label{eq:def-rb-ratio}
\end{align}
where $\ktminZ$, as mentioned, regulates the divergence.

The values of $q_\perp$ and $\ktminZ$ constitute the two free parameters of the model and are to be determined from 
a comparison with experimental data. From the construction, we expect their values 
to be approximately between $\LQCD$ and the perturbative cutoff $Q_0$ in the gluon PDF.
Using Eq.~(\ref{eq:def-rb-ratio}) in Eq.~(\ref{eq:prob-from-rb}) results in a probability
$P(\ktmin)$ for the effective colour screening that depends on the internal
kinematics of the system $X$.

We calculate the diffractive reduced cross section $\sigma_r^D(Q^2,\beta,\xP)$
within the same kinematic limits as applied by the experiment \cite{Aaron:2012ad}.
In addition, we adopt two different notions of the diffractive cross section.
The first definition $\sFWD(Q^2,\beta,\xP)$ is based on a forward remnant system
with a mass $M_Y < \unit[1.6]{GeV}$ and proton quantum numbers.
The other definition $\sLRG(Q^2,\beta,\xP)$ requires a large rapidity gap (LRG)
of two units in pseudorapidity.
This choice is potentially sensitive to the inner radiation structure of the
system $X$ because the LRG is defined in terms of pseudorapidity.

\subsection{Small-$x$ resummation via CCFM evolution}

The emissions in the CCFM evolution \cite{CCFM} are not strongly ordered in virtuality as they are by assumption
in the DGLAP evolution \cite{DGLAP}. Therefore, the parton in the hard interaction can no longer
be approximated as on-shell and instead an off-shell matrix element is used for the hard interaction.
Likewise, the branching gluons in the CCFM evolution are described by
an unintegrated gluon density function (UGDF).

In the CCFM evolution, one effectively resums the leading logarithms in the
energy splitting $1/z$ and $1/(1-z)$, as well as the leading logarithms in $Q^2$.
Experimental data on the DDIS cross section $\sigma^D(Q^2,\beta,\xP)$
covers a wide kinematic range where, in particular, one can have $M_X^2 \gg Q^2$.
Because of such potentially very different hard scales, we expect corrections from large
logarithms to become important in certain parts of the phase space.
Because of Eq.~(\ref{MXdef}) we expect the CCFM evolution to be better suited for the DDIS 
observables, at least, in the case of $\beta \ll 1$ when the large leading logs
\begin{eqnarray*}
\ln\frac{M_X^2}{Q^2} \simeq -\ln\beta \gg 1
\end{eqnarray*}
are properly treated.

\subsection{Soft divergences}

Because of the soft divergence in the first order QCD matrix element, the hard process
$\gamma g \to q\bar{q}$ will favour an uneven splitting of the energy between the quarks.
Specifically, if we define the fraction of energy taken by the quark as $z$,
then the matrix elements have soft divergences as $1/z$ and $1/(1-z)$
for $z\to 0$ or $z\to 1$, respectively. One aspect of the soft divergence is that it favors a rather 
large ratio between $x$ and the energy fraction $x_n$ of the parton entering the matrix element.
Specifically, in the matrix element $\gamma g \to q\bar{q}$ one can have a large ratio
$x/x_n$ without any additional radiation from the quark propagator.
Such a large $\ln 1/z$ is not accounted for in the initial state parton evolution.
In the case of the matrix elements $\gamma q \to q$ and $\gamma q \to q g$ plus
DGLAP evolution this phase space is in part taken into account,
but the DGLAP evolution does not resum potentially large $\ln x/x_n$.

Another aspect of the uneven splitting is that one of the quarks will be very forward in pseudorapidity
in the lab frame when $z$ is close to the divergence. In this case, the quark can have a large enough 
forward momentum $p_z$ to populate the gap region and the event therefore does not contribute 
to the diffractive cross section as defined in terms of a LRG in spite of having a leading proton.

The divergent behaviour is unphysical and is usually avoided with a cutoff.
Still, an inclusion of higher order effects may be important for the cross section
and also for the inner structure within the $X$ system. In particular, the energy of the $q\bar{q}$ 
system may be shared by additional gluons and therefore significantly reduce the rapidity range 
of the final $X$-system and therefore can have an influence on the LRG observable.

\begin{figure}[h]
\includegraphics[width=0.95\linewidth]{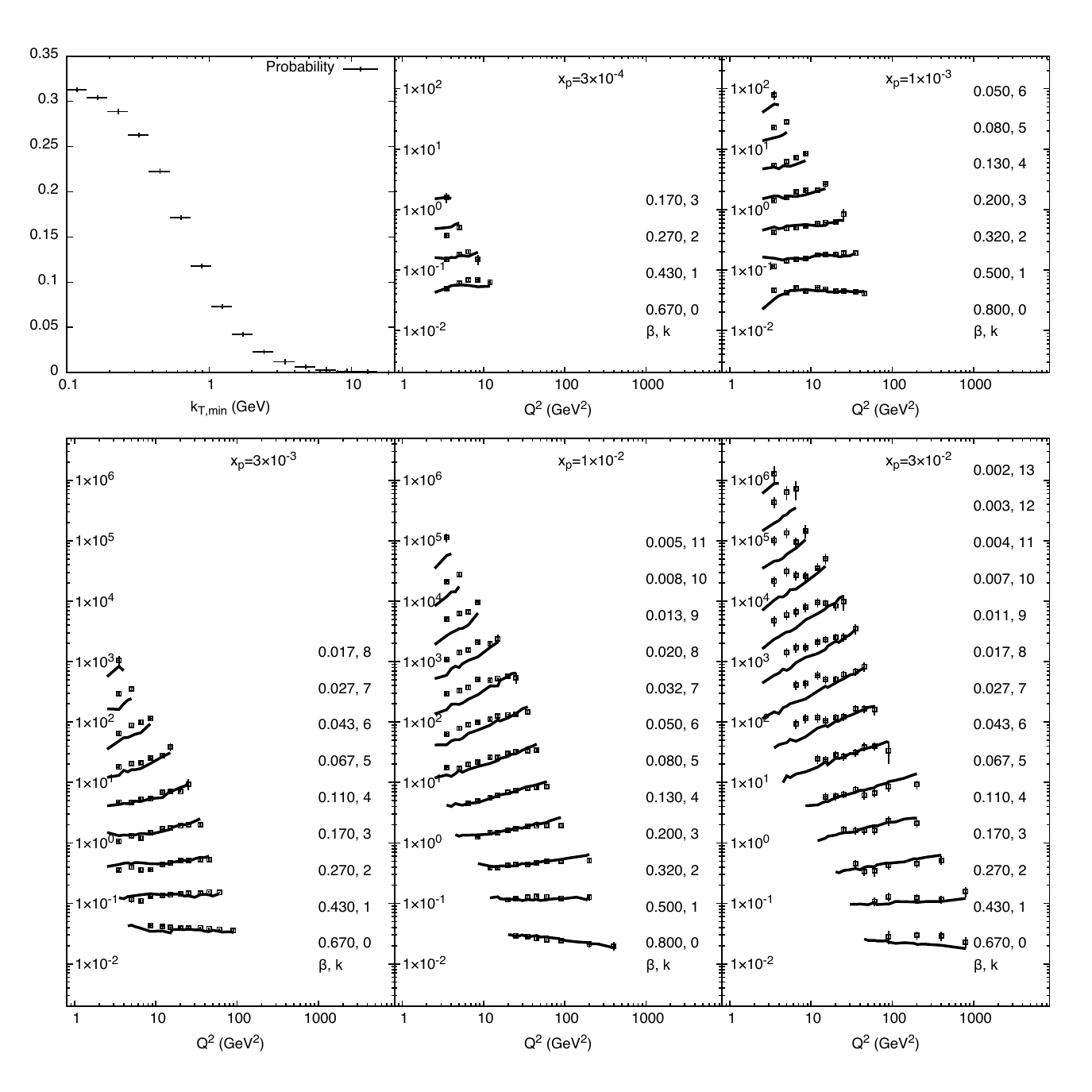}
\input{cascade_cutsnone_m01alphas70-params.inc}
\caption{
The reduced diffractive cross section $\sigma^D(Q^2,\beta,\xP)$ in comparison with
the H1 data \cite{Aaron:2012ad}.
The model prediction uses dynamic colour screening with the parameters \ppppK{} and \ppppQ{}
and the {\CASCADE} event generator with CCFM evolution.
$P(\ktmin)$ for the fitted parameters is shown in the upper-left corner.
Diffractive events in the model are defined as having a remnant system $Y$ with
proton quantum numbers and invariant mass $M_Y < \unit[1.6]{GeV}$.
Rows for different values of $\beta$ are offset by a factor $3^k$ as indicated on the figure.}
\label{fig:cascade_cutsnone_m01alphas70}
\end{figure}

\subsection{Details of the Monte Carlo implementation}
\label{sec:mc-impl}

In order to study the dynamic colour screening in more detail,
we interface the model with different Monte Carlo event generators. In particular, we employ
the program {\LEPTO} \cite{Ingelman:1996mq} which offers first order QED
and first order QCD matrix elements combined with DGLAP \cite{DGLAP} parton showering 
and collinear PDFs. As a second program we use {\CASCADE} \cite{CASCADE}
which offers $\gamma^* + g^* \to q\bar{q}$ with $k_\perp$-factorised off-shell matrix 
elements and CCFM evolution \cite{CCFM} which is intended to account for potentially large 
logarithms of incident momentum fractions of radiated partons.
The photon-gluon fusion matrix element $\gamma + g \to q\bar{q}$ illustrated
also in Fig.~\ref{fig:DDIS} is the dominant contribution to the diffractive DIS
cross section at small $x$. {\LEPTO} includes this process as a first order QCD matrix element,
as well as via a combination of the QED hard process $\gamma^* q \to q$ augmented by a
$g\to q\bar{q}$ DGLAP splitting. {\CASCADE} provides this process as an off-shell first order 
QCD matrix element, but not as a first order QED matrix element with parton splitting.

After generating events on parton level using matrix elements augmented
with initial and final state parton showers, we apply the colour screening model before any special 
treatment of the remnant. In particular, we do not allow any cluster fragmentation of systems 
with a small invariant mass because the dynamic screening will potentially change the colour topology 
of the event before the scale of hadronization is reached and therefore change the possible
outcomes of the fragmentation. This is understood as colour rescattering to happen on the scale 
between the perturbative cutoff at $\sim \unit[1]{GeV}$ and $\LQCD$ of hadronisation.

The remnant system is in a Monte Carlo program treated by a non-perturbative model.
For the case that the perturbative interaction resolves a gluon, which is the class
of events which potentially leads to diffraction, the remnant is usually split into a $(qq,q)$ pair
with a certain sharing of momenta. This splitting typically introduces a relative transverse 
momentum representing the Fermi motion in the bound state proton which is given by 
a Gaussian distribution with a width $\sim \LQCD$. This relative $k_\perp$ affects 
the later hadronisation which introduces an uncertainty for the prediction of a forward 
proton spectrum. In this work, we are interested in diffraction defined by a forward small-mass
system with proton quantum numbers or a large rapidity gap,
which is insensitive to whether the hadronisation model
maps the small-mass forward remnant state to a proton state or a resonance.

All plots for the diffractive cross section in this paper show the reduced cross section 
$\sigma_r$ which is related to the cross section via
\begin{align*}
\frac{d\sigma}{dQ^2\,d\beta\,d\xP} = \frac{4\pi\alpha_{em}^2}{\beta Q^4}
	\left( 1-y+\frac{y^2}{2} \right) \sigma_r(Q^2,\beta,\xP)
\end{align*}
with $y$ given as
\begin{equation*}
y = \frac{Q^2}{x (s-m_N^2)} \simeq \frac{Q^2}{xs}
\end{equation*}
with the negligible nucleon mass $m_N$.

\begin{figure}
\includegraphics[width=0.80\linewidth]{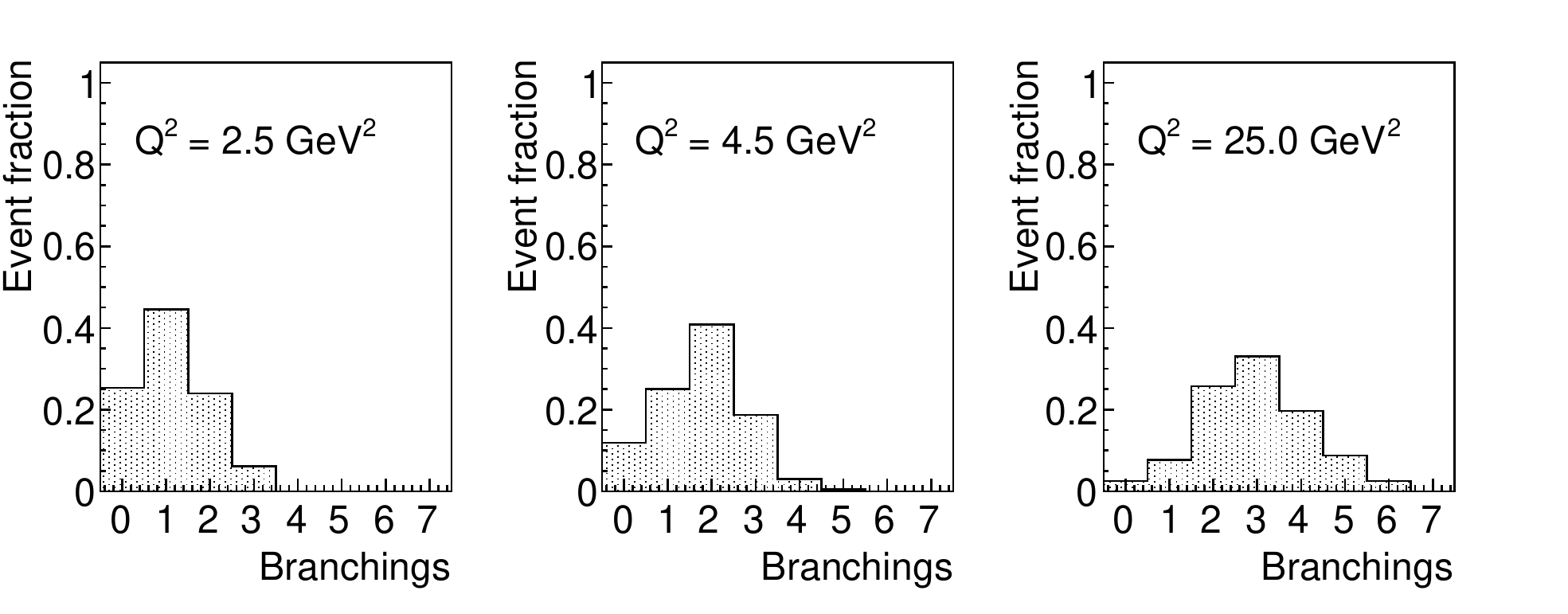}
\caption{
The distribution of the number of initial state radiation branchings in the CCFM evolution
at $\xP=3\times10^{-2}$, $\beta=0.017$ and for different values of $Q^2$.
As the hard scale increases, there is more phase space for initial state radiation available and we see more
gluon branchings. As the number of gluon branchings increases, a larger part of the full perturbative event
and specifically its $\beta$-value is described using CCFM evolution which leads to a more
accurate differential cross section.}
\label{fig:branchings-3em2-17em3}
\end{figure}

\begin{figure}
\includegraphics[width=0.95\linewidth]{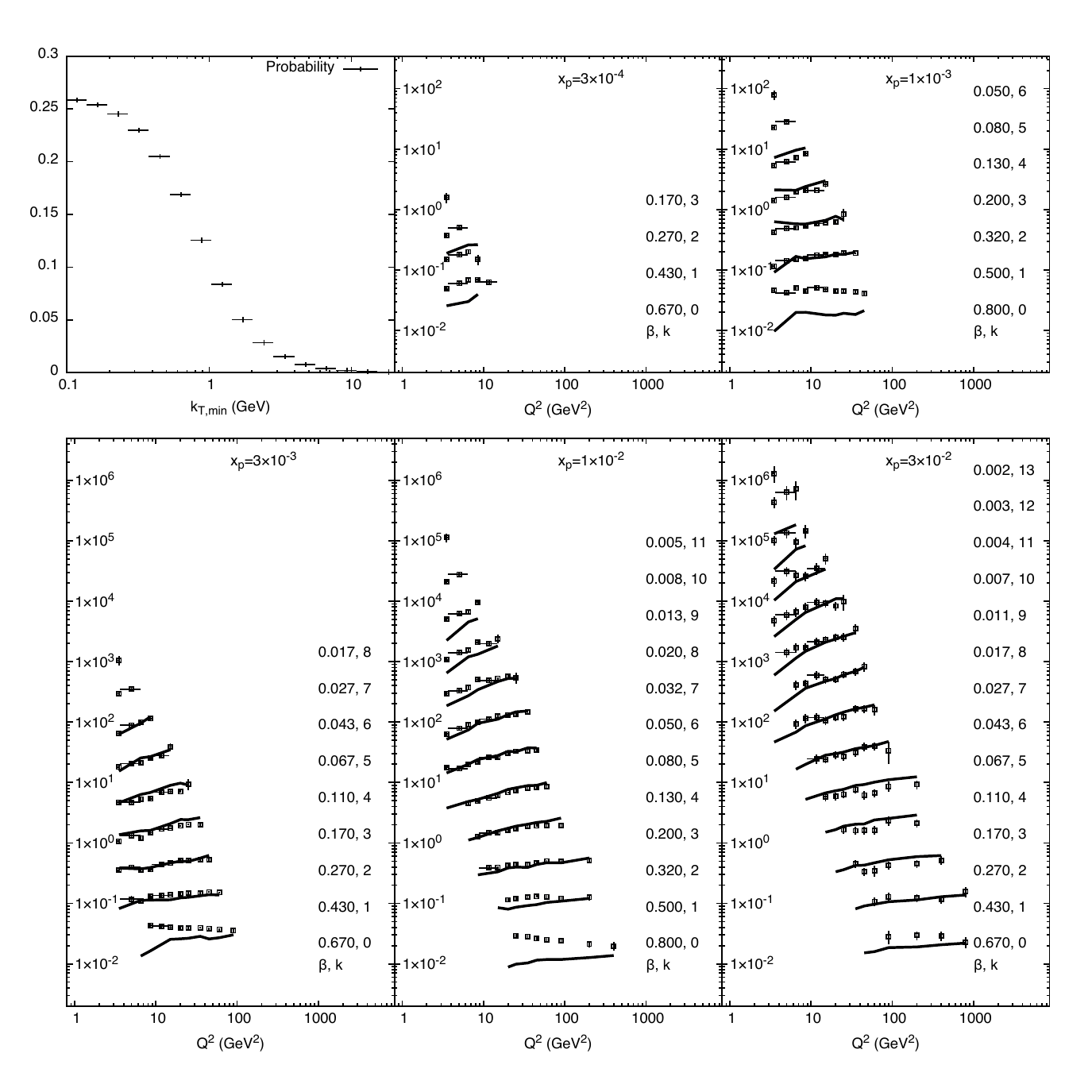}
\input{lepto_cutsnone-params.inc}
\caption{
The reduced diffractive cross section $\sigma^D(Q^2,\beta,\xP)$ as in
Fig.~\ref{fig:cascade_cutsnone_m01alphas70} but model results using the {\LEPTO} event 
generator based on matrix elements for all hard processes to order $\alpha_s$ and parton showers based on DGLAP $\log Q^2$ evolution. The parameters for the dynamic colour screening model obtained from the fit against data \cite{Aaron:2012ad} are \ppppK{} and \ppppQ{}.}
\label{fig:lepto_cutsnone}
\end{figure}

\begin{figure}
\includegraphics[width=0.95\linewidth]{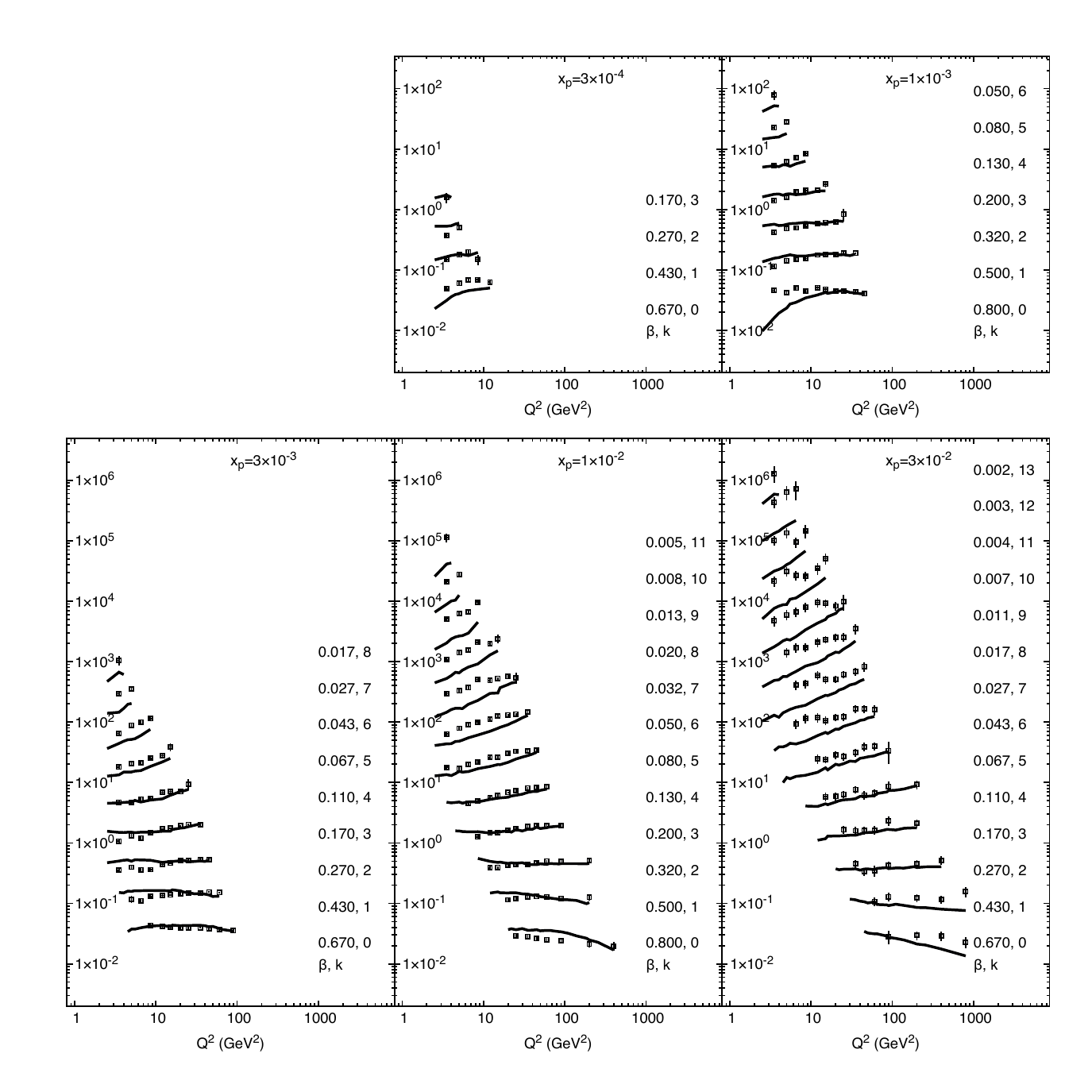}
\input{cascade_cutsnone_const-params.inc}
\caption{
As in Fig.~\ref{fig:cascade_cutsnone_m01alphas70} but with a constant probability
of $P=\parProb$ fitted from this data. We observe that especially at small $\beta$ the resulting 
diffractive cross section has a significantly different slope with respect to $Q^2$ and the description 
of data is not as good as in the case of the dynamic rescattering model.}
\label{fig:cascade_cutsnone_const}
\end{figure}

\section{Results}
\label{sec:res}

\subsection{Dynamic colour screening}

\input{cascade_cutsnone_m01alphas70-params.inc}
Fig.~\ref{fig:cascade_cutsnone_m01alphas70} shows the diffractive cross section
$\sFWD$ obtained with dynamic colour screening and the process $\gamma^* g^* \to q\bar{q}$
from the {\CASCADE} event generator with CCFM evolution. Events are selected according to the forward 
small-mass system prescription which requires a remnant system $Y$ with
proton quantum numbers and invariant mass $M_Y < \unit[1.6]{GeV}$.
The two parameters of the rescattering model are fitted and we obtain
\ppppK{} and \ppppQ{}. These values are physically reasonable in the sense that both are between
$\LQCD$ and $Q_0$, and that the typical transverse momentum scale $q_\perp$ of the proton background 
is smaller than the minimal transverse momentum scale $\ktminZ$ of the partonic $X$ system.

We note that there is an overall good agreement with experimental data
over a very wide region of the kinematical space. This agreement is remarkable 
because the model does not introduce specialised diffractive parton distributions,
but uses standard proton UGDFs as input and introduces only two new physically 
motivated parameters.

Nevertheless, we note that one specific kinematic region is not very well described, namely, 
where $\beta$ and $Q^2$ are both small and $\xP$ is large. The discrepancy develops for 
$\beta\lesssim 0.02$ and increases for decreasing $Q^2$ in the range of a few ${\rm GeV}^2$. 
A qualitative understanding of the problem can here be obtained from the principles of parton evolution via gluon radiation.
At small $\beta$, meaning large $M_X^2$ compared to $Q^2$, large logarithms of $1/\beta$ become important. 
On the other hand, at small $Q^2$ scales, the event is mainly described by the matrix element, whereas the shower 
activity is low, which leads to a relative damping of the cross section with respect to the data.
This lower radiative activity is illustrated in Fig.~\ref{fig:branchings-3em2-17em3} in terms of 
the number of branchings in the parton evolution for different $Q^2$ at a representative value of $\beta=0.017$.

At very small $\beta=x/\xP$ there is a very long evolution path from the initial gluon with momentum fraction $\xP$ to 
a much smaller Bjorken-$x$ at the quark-photon vertex. The {\CASCADE} event generator has the advantage to resum 
both $\log Q^2$ and $\log 1/x$ contributions in its CCFM-based treatment of initial-state gluon radiation off the incoming 
gluon. However, gluon emissions from the quark propagator between the $g\to q\bar{q}$ vertex and the quark-photon vertex 
({\it cf.} Fig.~\ref{fig:DDIS}b) can also generate large $\log 1/x$ contributions, in particular for the longest total evolution 
paths at very small $\beta$. This radiation from the quark propagator cannot be generated in {\CASCADE} since the CCFM 
equation only includes gluon radiation from gluons and not from quarks. The simulation process starts by using the hard matrix 
element for $\gamma^*g^* \to q\bar q$, and then the initial gluon is evolved down to the starting scale of the proton state. 
Thus, at very small $\beta$ in particular, there can be a substantial phase space available for gluon radiation off the quark 
propagator connecting to the virtual photon, which is not taken into account.

Gluon emission from the quark propagator can be taken into account by instead using the {\LEPTO} Monte Carlo generator. 
It includes not only the hard matrix element for $\gamma^*g \to q\bar q$ but also the other first order QCD process 
$\gamma^* q \to qg$, where the gluon can be emitted from the incoming or outgoing quark, as well as the zeroth-order 
process $\gamma^* q \to q$. For both these processes, additional gluon radiation from the initial quark may occur through 
the initial-state parton showering, which in {\LEPTO} is generated through conventional DGLAP evolution in $\log Q^2$, 
but without a resummation of $\log 1/x$ contributions.

Fig.~\ref{fig:lepto_cutsnone} shows the results of using {\LEPTO} with the same dynamic colour screening model, 
with fitted parameter values \ppppK{} and \ppppQ{} of expected magnitudes. The description of data is good in 
the inner region of the covered kinematic space, but with substantial discrepancies at small $\xP$ and very large $\beta$. 
In the discussed problematic region, however, the agreement with data is better than that for {\CASCADE} at large $\xP$, 
small $Q^2$ and small $\beta$ around 0.01, although not at the very smallest $\beta\lesssim 0.004$. This indicates that, 
as discussed, emissions from the quark propagator is of some importance, but accounting for large $\log 1/x$ contributions 
is also needed.

No presently available event generator include all the mentioned effects that seems necessary in order to describe the diffractive 
cross-section over the entire kinematic region. The indication is, however, that combining all available theoretical formalisms 
in perturbative QCD, i.e. matrix elements and parton showers including $\log Q^2$ and $\log 1/x$ resummations, with 
the dynamical colour screening model presented here should provide a working description of the observed diffractive 
deep inelastic scattering process.

\subsection{Colour screening probability}

\begin{figure}
\begin{minipage}{0.40\linewidth}
\includegraphics[width=0.99\linewidth]{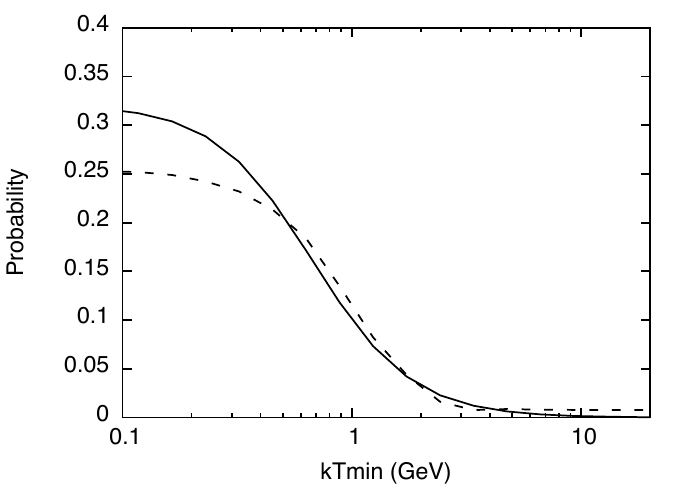}
\end{minipage}
\caption{
Dashed line: The fit of a free-form probability function $P(\ktmin)$ constrained only by the
requirement to be in the physically sensible range $[0,1]$ and to be fairly smooth.
Solid line: The probability from the fit of the colour screening model as used in
Fig.~\ref{fig:cascade_cutsnone_m01alphas70} (upper left corner).
We note that both methods result in very similar functions for $P(\ktmin)$.}
\label{fig:prob-casc-M01-alphas70}
\end{figure}

\input{cascade_cutsnone_const-params.inc}
It is interesting to compare the dynamic colour screening model with the results from
having a fixed colour screening probability while keeping all other parameters equal.
Fig.~\ref{fig:cascade_cutsnone_const} shows the diffractive cross section with $P=\parProb$,
obtained by a fit to data. We note that the constant probability results in a significantly worse
description of the data. The overall normalisation as well as the shape of $\sigma^D(Q^2,\beta,\xP)$
with respect to $Q^2$ is better described by the dynamic screening model.

In the results from the dynamic model of Fig.~\ref{fig:cascade_cutsnone_m01alphas70},
we have fitted the two parameters $\ktminZ$ and $q_\perp$ of the dynamic screening model to data.
The parameters determine the overall normalisation and essentially the position
of the slope where the screening probability $P(\ktmin)$ falls off to zero.
On the other hand, the general shape of this probability is given by the underlying
model itself. It is interesting though to investigate what form of $P(\ktmin)$ would result in a good fit
without assuming an underlying model. To this end, we fit a mapping $\ktmin \to P$ which is only constrained
by the fact that it should lie in the physically sensible range $[0,\,1]$
and that it should be reasonably smooth. The result of such a fit is shown in
Fig.~\ref{fig:prob-casc-M01-alphas70}. We observe that even though
we did not place any particular constraint on the functional form,
one obtains essentially the same result as in the dynamic screening model
and therefore support to its QCD-basis.

\subsection{Dependence on the gluon density}

\begin{figure}
\begin{minipage}{0.49\linewidth}
\includegraphics[width=0.7\linewidth]{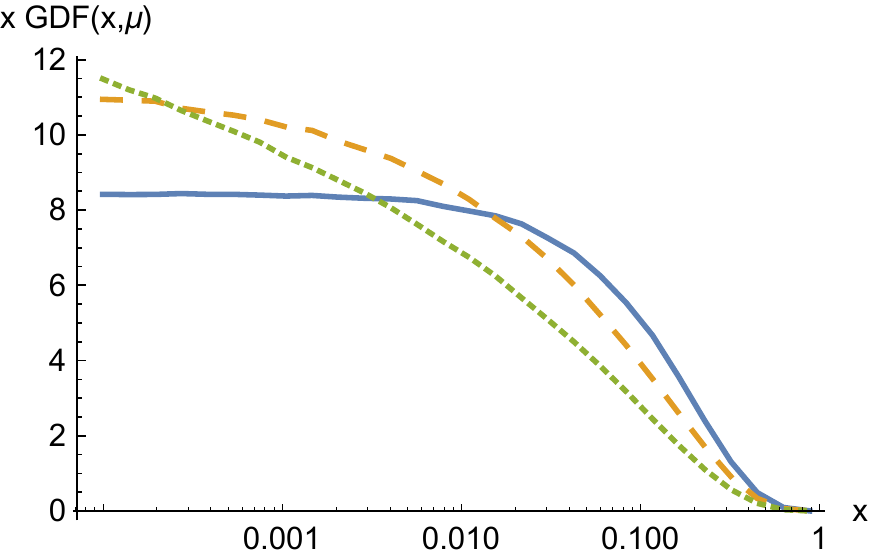}
\end{minipage}
\begin{minipage}{0.49\linewidth}
\includegraphics[width=0.7\linewidth]{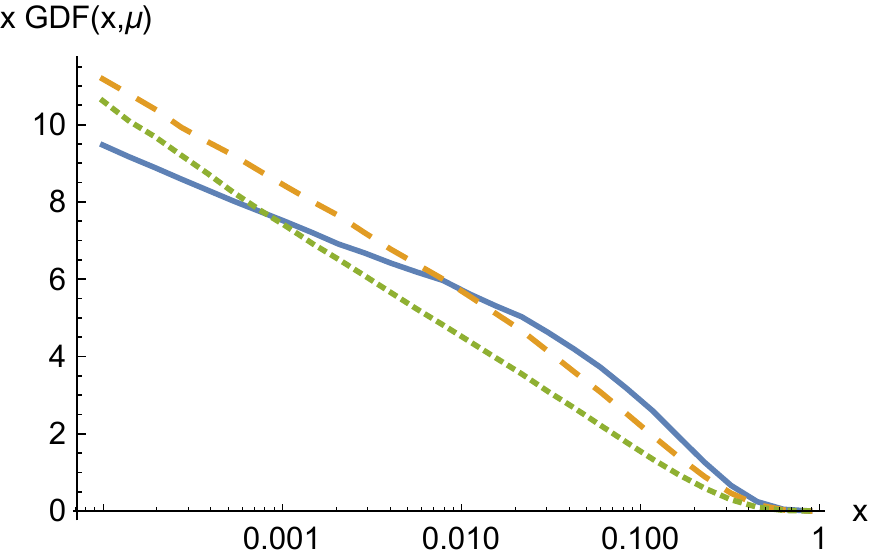}
\end{minipage}
\caption{
The gluon density \textsc{A0} from {\CASCADE} \cite{CASCADE} integrated over $k_\perp$
at the scales 1, 4 and $\unit[8]{GeV}$ for the solid, dashed and dotted lines, 
respectively, is shown in the left panel. The gluon density \textsc{A1} for the 
same scales is shown in the right panel. The distribution starts out with a steeper 
slope at small scales and influences the observable as shown in 
Fig.~\ref{fig:cascade_gdf1013_cutsnone_m01alphas70}.}
\label{fig:gdf-plot-1010}
\label{fig:gdf-plot-1013}
\end{figure}
\begin{figure}
\includegraphics[width=0.95\linewidth]{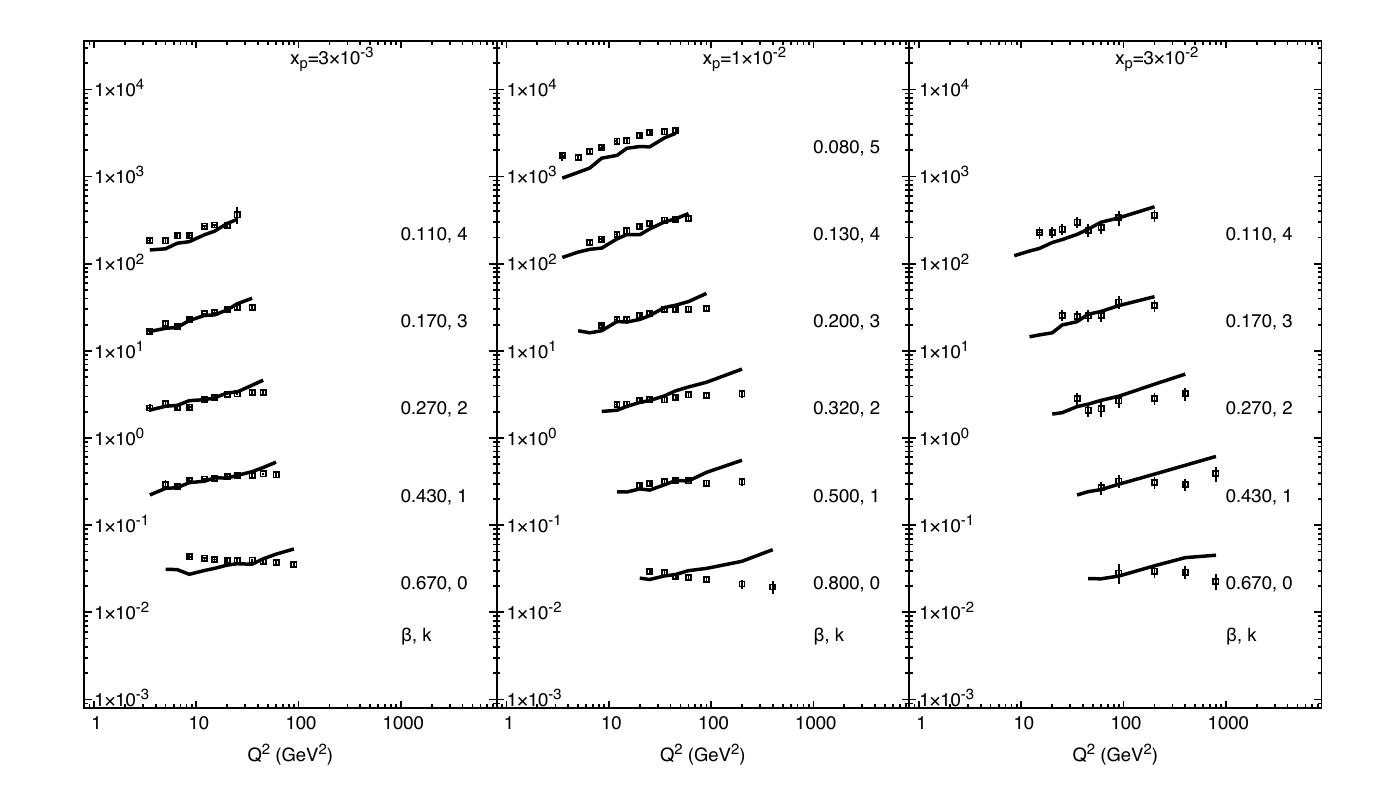}
\caption{
As in Fig.~\ref{fig:cascade_cutsnone_m01alphas70} but using a gluon density 
$xA(x,k_\perp,\mu)$ that increases stronger towards
low-$x$ already at the starting scale (\textsc{A0} in \cite{CASCADE}).
Only the interesting subset of the kinematic plane is shown.
By comparison with Fig.~\ref{fig:cascade_cutsnone_m01alphas70} it is seen that
$\sigma^D(Q^2,\beta,\xP)$ is at large $\beta$ and large $\xP$ sensitive
to the shape of the gluon density at the starting scale.}
\label{fig:cascade_gdf1013_cutsnone_m01alphas70}
\end{figure}

The result in Fig.~\ref{fig:cascade_cutsnone_m01alphas70}
is obtained using the unintegrated gluon density $xA(x,k_\perp^2,\mu)$ illustrated
in Fig.~\ref{fig:gdf-plot-1010} (left). This density starts out flat at a low scale $\mu$ which 
can be compared with the $1/\xP$ behavior of the Pomeron flux in Regge-based models.
The diffractive cross section in our model is sensitive to the slope especially in the kinematic 
region where $\beta$ is large. We can compare the main result in Fig.~\ref{fig:cascade_cutsnone_m01alphas70}
with the result obtained by using a parton density which has a stronger increase towards
small $x$ already at low scales, shown in Fig.~\ref{fig:gdf-plot-1010} (right).
The corresponding $\sFWD$ is shown in Fig.~\ref{fig:cascade_gdf1013_cutsnone_m01alphas70}.
We note that especially the dependence of the cross section on $Q^2$ is sensitive
to the gluon density $xA(x,k_\perp^2,\mu)$ at small scales $\mu$.
By including diffractive data into the fit of a gluon density,
this dependence could be used to further constrain
the shape of the gluon distribution at low scales.

\subsection{Diffractive cross section with a large rapidity gap}

\begin{figure}
\includegraphics[width=0.95\linewidth]{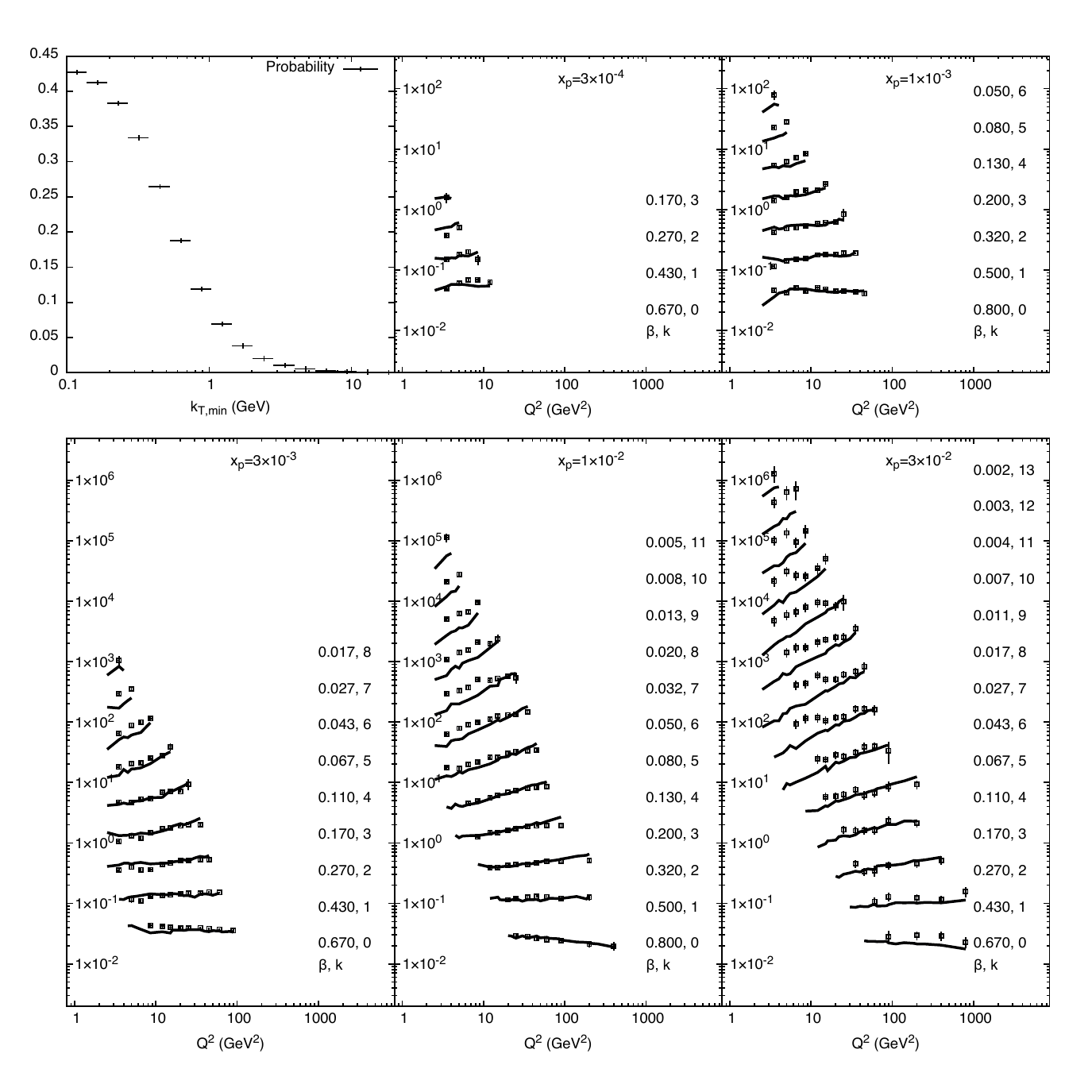}
\input{cascade_LRG40T60_m01alphas70-params.inc}
\caption{
As in Fig.~\ref{fig:cascade_cutsnone_m01alphas70} but showing
$\sLRG$ where the diffractive cross section is defined in terms of
a large rapidity gap between $\eta_{min}=4.0$ and $\eta_{max}=6.0$ 
in pseudorapidity.}
\label{fig:cascade_LRG40T60_m01alphas70}
\end{figure}
\begin{figure}
\begin{minipage}{0.495\linewidth}
\includegraphics[width=0.999\linewidth]{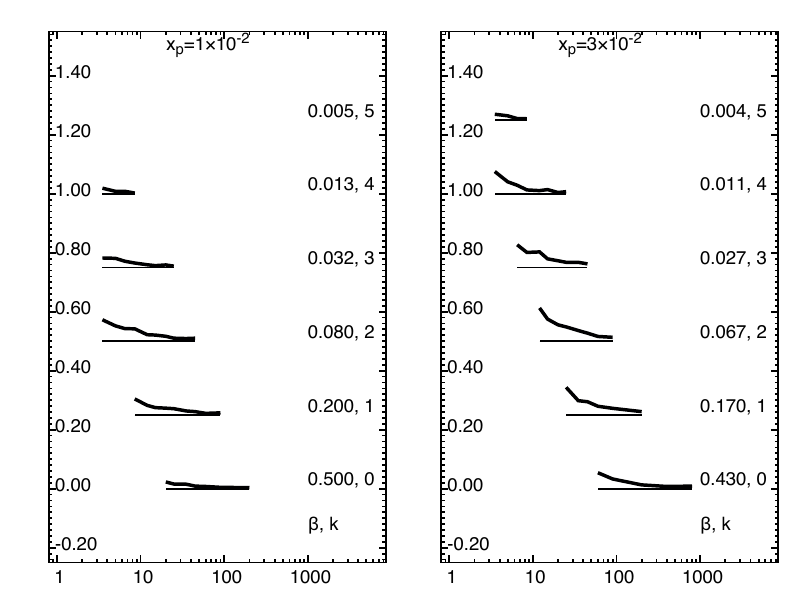}
\end{minipage}
\begin{minipage}{0.495\linewidth}
\includegraphics[width=0.999\linewidth]{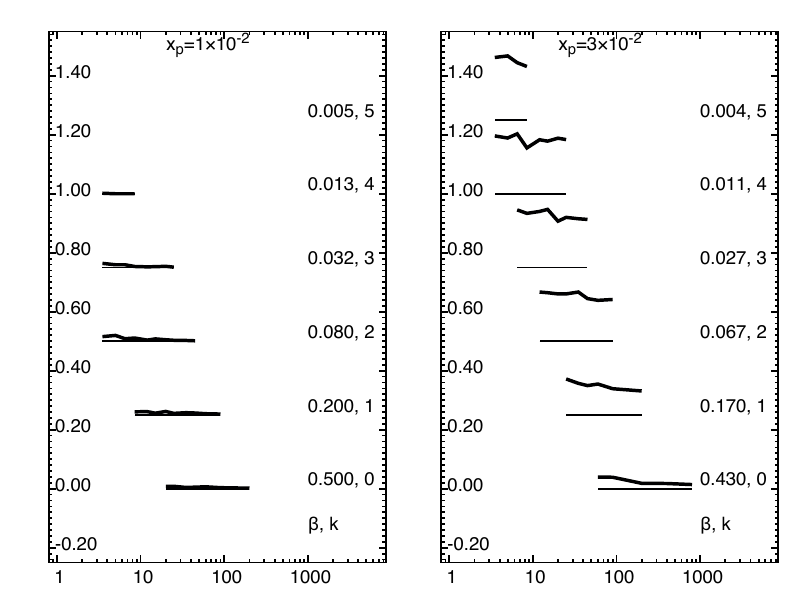}
\end{minipage}
\caption{
The fraction of events with a parton level quark with a pseudorapidity in the gap 
region is shown in the two leftmost plots. The 0\% level within the $Q^2$ range 
available in the experimental data is indicated by a horizontal thinner line for each row in $\beta$.
The fraction of events with a gluon in the LRG region is shown in the two rightmost plots.}
\label{fig:lrg-parton-veto}
\end{figure}

Fig.~\ref{fig:cascade_LRG40T60_m01alphas70} shows the diffractive cross section
$\sigma^D_\mathrm{LRG}(Q^2,\beta,\xP)$ as defined by the presence of a large rapidity gap
in the range $4 \le \eta \le 6$ in pseudorapidity. We note the close similarity to Fig.~\ref{fig:cascade_cutsnone_m01alphas70}.
When requiring a minimum gap size $\Delta\eta$ in the event, the non-diffractive cross-section is exponentially suppressed by $\Delta\eta$.
While the presence of a LRG of size $\Delta\eta$ in the final state is a clear indication of diffraction, the exact location of the gap is sensitive 
to the description of the kinematics of the central system $X$, and $\sigma^D_\mathrm{LRG}(Q^2,\beta,\xP)$ depends therefore additionally 
on the treatment of dynamics within the $X$-system. The dynamic colour screening model describes the probability of a colour screened 
interaction, but the additional soft exchanges do not change the kinematics of final state partons. We know on the other hand that our 
description in terms of matrix elements plus showering does not cover the full phase space. Especially, the lack of additional gluon radiation 
from the quark propagator in the matrix element $\gamma^* + g^* \to q\bar{q}$ could be the reason why this $q\bar{q}$-system extends 
up to too high rapidity and can hadronize into the gap region. We therefore keep the gap size $\Delta\eta = 2$ as in the data, but shift  
its starting value from 3.3 in data to 4. This modification is irrelevant for our result at $\xP < 0.01$, but noticeable at $\xP = 0.01$
where it causes at most a correction of $30\%$ at small $\beta$ and small $Q^2$, but becomes important at $\xP = 0.03$.

We expect $\sigma^D_\mathrm{LRG}(Q^2,\beta,\xP)$ to be sensitive to higher order corrections especially close to the soft divergence $1/z$
where partons from the matrix element cause activity in the LRG region. This is illustrated in Fig.~\ref{fig:lrg-parton-veto} where the fraction 
of events having a $q$ or $g$ with momentum vector into the LRG is shown. We note that the fraction of quarks in the LRG increases towards low $Q^2$ 
which can be understood from the correlation between $Q^2$ and the transverse momentum. On the other hand, the fraction of gluons in the LRG depends
more strongly on $\xP$ and weaker on $Q^2$. This can be understood from the fact that gluons arise from the parton shower
in contrast to the quarks which are defined by the matrix element.

At the leading order of our computation, we note that the result is only mildly sensitive to the cuts employed to regulate the $1/z$ divergence.
Also, a comparison with massive matrix elements at leading order suggests that typical quark masses lead effectively to the same cuts on the
energy sharing variable $z$ as used in our results. On the other hand, higher order corrections could significantly alter the internal
event structure, especially in the region of the phase space where it is likely to have a large step $z$ between $x$ and $x_n$.
There, an additional final state gluon could modify the extension of the $X$ system in rapidity. An improved parton evolution based 
on CCFM with additional splittings $g \to qq$ and $q \to qg$ could therefore in principle improve the description further because
the diffractive process at a low scale could be described with a QED matrix element and low $x$ resummed parton evolution.
Similarly, $\sigma^D_\mathrm{LRG}(Q^2,\beta,\xP)$ is also sensitive to the distribution of the transverse momenta and 
energy splitting in the parton shower.

\section{Summary and conclusions}
\label{sec:conclusions}

We have developed the probability for dynamic colour screening in DIS in a way
that can be used in Monte Carlo event generators and applied it with {\CASCADE}
and {\LEPTO}. The resulting model predicts the diffractive DIS cross section based on
perturbative QCD matrix elements and standard inclusive parton densities in
both collinear and $k_\perp$-factorisation approaches.
This facilitates practical applications of previously obtained theoretical
derivations of the amplitude for colour screening through semi-soft multiple
gluon exchanges calculated in the eikonal approximation to all orders in
perturbative QCD. The basic formalism gives a theoretical understanding why the soft colour interaction (SCI) model has been
phenomenologically successful, but goes beyond that model by leading to a
colour screening probability that depends on the dynamics of the perturbative QCD parton
dynamics. This dynamical screening probability exhibits a saturated behaviour at small
transverse momenta of the emerging parton system as well as colour transparency
at large transverse momentum.

The Monte Carlo model has only two, physically motivated parameters.
Their values are obtained by fitting the HERA diffractive cross section and
found to be of the expected magnitude.
The model successfully describes the data over a large kinematic range,
significantly better than with a constant screening probability.
Interestingly, a fit of a free-form probability function results in the same
shape as in our model, and hence gives support for our account of the basic QCD
dynamics of relevance. It is noteworthy that we have not introduced any diffractive 
parton distribution functions to describe soft dynamics through unknown functions 
fitted to data. This explicitly demonstrates that diffractive and inclusive scattering 
can be described by the same basic QCD processes when explicit account is taken of 
color degrees of freedom in gluon exchanges at scales below normal cut-offs of $\sim 1$ GeV 
of hard, perturbative QCD processes and $\Lambda_{\rm QCD}$ of the hadronisation phase transition.  

In some kinematic regions there are two very different scales present, namely
the invariant mass of the diffractive system $M_X^2$ and the photon virtuality
$Q^2$.  This calls for a resummation of large logarithms $\log M_X^2 / Q^2$, or
equivalently $\log 1/\beta$.
To address this issue, we take the cross section in the $k_\perp$-factorisation
approach from off-shell matrix elements and unintegrated gluon densities
together with the CCFM evolution which provides a resummation of leading
logarithms in $1/x$.
We show that this significantly improves the description of the diffractive
HERA data at $\beta\ll 1$ corresponding to $M_X^2 \gg Q^2$.
Nevertheless, there are residual deviations in the region where both $\beta$
and $Q^2$ are at their lowest values.
This may be attributed to the extreme region of very small $\beta$ when the quark
propagator connected to the virtual photon have a significant phase space
available for gluon radiation, which is not accounted for;
neither in the leading order matrix element for $\gamma^\star g^\star \to q\bar{q}$
nor in the CCFM evolution that does not include the $q \to qg$ splitting.
The DGLAP evolution does include this and also shows a slightly better result in
this case, but is still not sufficient since here the effects of large
$\log1/x$ is not included.

To conclude and connect to the discussion in the Introduction, our study has
shown that the phenomenon of diffractive deep inelastic scattering can be
described using a basic QCD-framework. The hard subprocess is treated in the same way
as for non-diffractive events but a colour screening process occurs as a result of
multiple gluon exchanges that are resummed to all orders. Significant
deviations from data occur in a special kinematic region, where potentially
large logarithmic corrections are not yet fully included in available evolution
equations for gluon radiation. Still, the overall results show that gluonic
colour screening in QCD is a viable approach to understand
diffraction.

{\bf Acknowledgments}
This work was supported by the Swedish Research Council under 
contracts 621-2011-5107 and 621-2013-4287.


\end{document}

%% file: lepto_cutsnone-params.inc
\providecommand\ppppQ{}
\renewcommand{\ppppQ}{$q_\perp=\unit[0.66]{GeV}$}
\providecommand\ppppK{}
\renewcommand{\ppppK}{$\ktminZ=\unit[0.89]{GeV}$}
\providecommand\parChiSq{}
\renewcommand{\parChiSq}{2688}

%% file: cascade_cutsnone_m01alphas70-params.inc
\providecommand\ppppQ{}
\renewcommand{\ppppQ}{$q_\perp=\unit[0.58]{GeV}$}
\providecommand\ppppK{}
\renewcommand{\ppppK}{$\ktminZ=\unit[0.72]{GeV}$}
\providecommand\parChiSq{}
\renewcommand{\parChiSq}{666}

%% file: cascade_cutsnone_const-params.inc
\providecommand\parProb{}
\renewcommand{\parProb}{0.103}
\providecommand\parChiSq{}
\renewcommand{\parChiSq}{1898}

%% file: cascade_LRG40T60_m01alphas70-params.inc
\providecommand\ppppQ{}
\renewcommand{\ppppQ}{$q_\perp=\unit[0.54]{GeV}$}
\providecommand\ppppK{}
\renewcommand{\ppppK}{$\ktminZ=\unit[0.57]{GeV}$}
\providecommand\parChiSq{}
\renewcommand{\parChiSq}{685}